\documentclass[12pt]{article}
\usepackage{amsmath,amsfonts,color}
\usepackage[numbers,sort&compress]{natbib}
\usepackage{times}
\usepackage[left=2.54cm,top=2.54cm,right=2.54cm,bottom=2.54cm,bindingoffset=0.0cm]{geometry}
\usepackage{setspace}
\usepackage{mdframed}

\usepackage{graphicx}
\usepackage{amssymb}
\usepackage{vaucanson-g}

\usepackage{imakeidx}
\makeindex

\begin{document}
\parskip 0pt

\title{Major Transitions in Political Order}
\author{Simon DeDeo\footnote{Center for Complex Networks and Systems Research, Department of Informatics, Indiana University, 919 E 10th St, Bloomington, IN 47408; Program in Cognitive Science, Indiana University, 1900 E 10th St, Bloomington, IN 47406; Ostrom Workshop in Political Theory and Policy Analysis, 513 N Park Avenue, Bloomington, IN 47408; Santa Fe Institute, 1399 Hyde Park Road, Santa Fe, NM 87501, USA. {\tt simon@santafe.edu}}}

\maketitle
\begin{abstract}
\noindent
We present three major transitions that occur on the way to the elaborate and diverse societies of the modern era. Our account links the worlds of social animals such as pigtail macaques and monk parakeets to examples from human history, including 18th Century London and the contemporary online phenomenon of Wikipedia. From the first awareness and use of group-level social facts to the emergence of norms and their self-assembly into normative bundles, each transition represents a new relationship between the individual and the group. At the center of this relationship is the use of coarse-grained information gained via lossy compression. The role of top-down causation in the origin of society parallels that conjectured to occur in the origin and evolution of life itself.
\end{abstract}

\noindent {\it [T]hey then threw me upon the bed, and one of them (I think it was Mary Smith) kneeled on my breast, and with one hand held my throat; Mary Junque felt for my money; by my struggling about, they did not get it at that time; then they called another woman in \textellipsis when she came in, they said cut him! cut him!} --- evidence of Benjamin Leethorp in the trial of Mary Junque and Mary Smith for grand larceny, Old Bailey Criminal Court, London, England; 4 April 1779~\cite{obo_site}\index{Old Bailey}

\vspace{0.5cm}

\noindent
Unless we are historians, the 18th Century world of Junque, Smith and Leethorp is almost impossible to imagine. In stealing from Leethorp, the two women put themselves at risk not only of imprisonment, but of indentured servitude in the colonies and even death. Leethorp, for his part, begins his evidence by explaining to the jury how he was seeking a different brothel than the one in which he was throttled, stripped, and robbed. Junque and Smith were without benefit of legal counsel and Smith's witnesses, unaware of the trial date, did not appear. The court condemned them to branding and a year's imprisonment in less than five hundred words. The indictment, formally for a non-violent offence, was one of hundreds of its kind that decade marked by assault, knives, and (sometimes) freely flowing blood.

In the risks they ran and the things they were ashamed of, the minds of the three are alien to us; in its casual violence, so was the society that enclosed them. Yet this world, gradually, continuously, evolved into one far less tolerant of violence and yet far more protective of an individual's rights---into the world, in other words, of most readers of this volume. How witnesses, victims, and defendants spoke about both facts and norms in the law courts of London shifted, decade by decade, over the course of a hundred and fifty years~\cite{obo}. This shift in speech paralleled a similar decline in how people behaved towards each other on the street, as the state came, increasingly, to manage its monopoly on violence---part of what is known as the ``civilizing process''~\cite{elias}.\index{civilizing process}

These changes took place in the decentralized common-law courts, among hundreds of thousands of interacting victims and defendants. Acts of Parliament, sensational crimes, the invention of the criminal defence lawyer---these changed the courts, but in the moments of their introduction showed little effect on the slow changes in the speech and practices themselves. We are predisposed to see the introduction of a law as identical to the recognition and enforcement of the moral sentiments it invokes. Yet it is, in the final analysis, individuals who constitute a social world. Laws and formal practices may be created by a small group that can unilaterally enforce its will, but they often lag behind the conditions they ratify; when laws do appear, they have unpredictable effects on the minds of the people they concern~\cite{gneezy2000fine,bowles2008policies}.

Evidence from the quantitative behavioral and social sciences accumulates daily for the existence of a complex relationship between individual minds and the persistent social worlds they create\index{individual and group, relation between}. Over decades of development, writers collectively nucleate new styles of prose on the periphery of the generation that came before, perceiving the patterns of the past and struggling with their influence~\cite{krakauer}. French revolutionaries borrow words such as {\it contract}, {\it rights} and {\it the people} from Enlightenment philosophers to both signal and make possible their shifting political alliances~\cite{baker1990inventing}; these same words appear, hundreds of years later, as signals in the House and Senate of 21st Century America~\cite{rion}. Pre-Hispanic Mexico and 21st Century Europe have similar patterns in the distribution of city sizes, outputs, and infrastructure, showing how widely-varying cultures find similar solutions to the management of social contact over more than three millennia~\cite{ortman2014pre}.

Such phenomena are often called political, but {\it homo sapiens} is not the only political animal. As we will show, increasing evidence from the behavioral sciences shows that social animals such as pigtailed macaques and monk parakeets interact not only with each other, but with the creations of their society as a whole. As we approach our own branch on the evolutionary tree we find a sequence of transitions in the nature of the relationship between the individual and the group: individuals come to know coarse-grained facts about their social worlds; they gain the ability to reason normatively, from a collective ought; they gather their norms into self-reinforcing bundles. New research provides a quantitative window onto the distinct and traceable imprints each of these transitions leaves on the logic of society.

In their book \emph{Major Transitions in Evolution}, John Maynard Smith and E\"{o}rs Szathm\'{a}ry~\cite{smith1997major} argued that leaps in complexity over evolutionary time were driven by innovations in how information is stored and transmitted. Our social feedback hypothesis\index{social feedback hypothesis} extends their argument to account for the major transitions in political order. We argue that these later transitions are driven by innovations in how information is \emph{processed}.\footnote{We do not, however, describe the evolutionary pressures that might drive the creation of these novel abilities; most notably, the collective action problem~\cite{blanton2007collective,carballo2014cooperation}, whose study has formed the basis of fruitful contact between the anthropological and political sciences.} 

Our attention to information processing focuses in particular on the summary of large numbers of individual-level facts to produce coarse-grained representations of the world. Understanding what coarse-graining is, and how it works, is essential. We begin there.

\section{Coarse-graining the Material World}\index{coarse-graining}

To build a scientific account of the origins and major transitions in political order, we turn first to a question at the heart of 20th Century physics: what is the charge of the electron? This apparently simple problem of measurement is far more subtle that it appears, and its resolution was a major advance with unexpected implications. 

With the classical theory of electromagnetism---the one taught in high school---it is simple to devise any number of experiments that can measure the electron charge, which appears constant no matter how it is studied. But the extensions of electromagnetism to the quantum domain are far less tractable: depending on the calculations one does, the apparent charge varies and can even, when the mathematics are worked out, diverge.\index{electromagnetism}

In response to this unacceptable state of affairs, physicists considered the idea that the charge of the electron might vary depending upon the scale---literally, the physical size---on which the experiment is done. Rather than construct an explicit, mechanistic account of the electron's substance, they developed a theory that described the dynamics of a smoothed-out version of the electromagnetic fields it creates. The averages of fields on centimeter scales obey one set of laws, the averages on nanometer scales, another. This means that, as you retain information about smaller and smaller distances, the implied properties of the electron shift rather than stabilize.\footnote{The process by which these properties changed was, for historical reasons, given the name ``renormalization''; see Ref.~\cite{kadanoffstatistical} for a simple introduction, and Ref.~\cite{caobook} for extended discussion.}\index{renormalization} 

Electrodynamics is just one example of how physicists built a theory not on a detailed account of underlying mechanisms, but on the rules obeyed by averaging their effects. To do this averaging in the case of electromagnetism, physicists were naturally drawn to the idea of a spatial average. When mechanisms are local---when a point $X$ can influence a point $Y$ only via intermediate points between $X$ and $Y$---one can retain a great deal of predictive power by averaging together points that are physically nearby. Because of how influence propagates, it makes little sense to average together two distant points; conversely, we can build a reliable, if only partial, theory from considering the interactions between neighborhoods.

A simple example of this spatial coarse-graining is provided by cellular automata.\index{cellular automata} These discrete, spatially-organized systems are governed by a deterministic local mechanism. The state of any point in the system is determined by the neighbors of that point at the time-step before. If we ``squint''---\emph{i.e.}, if we blur the system, averaging nearby points and reducing the resolution---the objects of the new, coarse-grained system will obey a different set of laws. 

We have lost information when we summarized, or compressed, or shortened the representation. This lossy compression\index{lossy compression} means that some of the information necessary to predict the fine-grained evolution has been dropped; in general, this loss of predictivity will affect the coarse-grained level as well, making a system that is fundamentally deterministic appears to follow probabilistic laws.

\begin{figure*}
\centering
\includegraphics[width=0.5\textwidth]{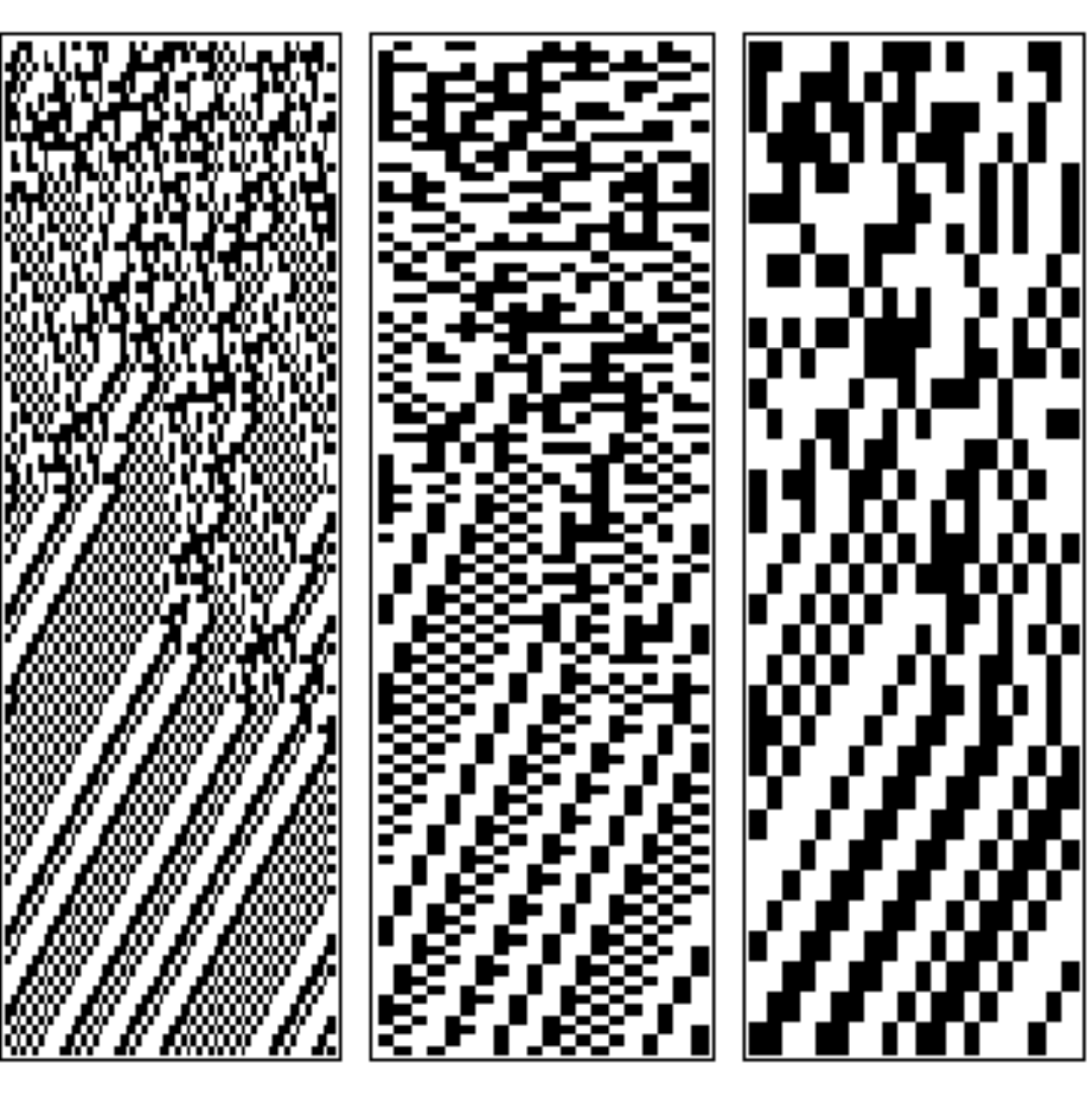}
\caption{Coarse-graining a one-dimensional cellular automata. Left panel: one hundred iterations of the $r=2, W=360A96F9$ rule, with random initial conditions; middle panel: the same run, coarse-grained by Kadanoff decimation along the spatial axis. While perceptual and memory costs are reduced by a factor of three, the coarse-grained system becomes harder to predict and deterministic rules become probabilistic in nature. Right panel: the same run, coarse-grained in both time and space. The system is now simplified by a factor of nine; we preserve approximate relationships, and rough, probabilistic logics of evolution.\label{ca}}
\end{figure*}

An example is shown in Fig.~\ref{ca}; we begin with the exact solution, down at the mechanism scale (left panel). If we coarse-grain in space (middle panel) or both space and time (right panel), we have fewer blocks to keep track of while still preserving some of the gross features of the system (such as the transition, in this figure, to diagonal order around the mid-way point). In dropping fine-grained complexity, however, our new logic becomes probabilistic, not deterministic. We have gained simplicity at the cost of predictive accuracy. The so-called critical points of this phenomenon, for the case of cellular automata, have been investigated in elegant detail by Edlund and Jacobi~\cite{edlund2010renormalization}.

Spatial coarse-graining is not the only way to simplify a system, and in many cases may not be appropriate. When we move from the physical to the biological or social sciences we find systems that are fundamentally long-range in nature, or have mechanisms that tie together distant locations. Averaging nearby points might simplify the system, but destroy any possibility of finding a reasonable model to relate these coarse-grained states. In a cell, for example, a fragment of RNA should not be averaged with nearby molecules to describe a cell in terms of the local density of its cytoplasm; better descriptions might summarize the counts of different RNA sequences within the cell, even when they occur at large spatial separations. A lossy compression is to be evaluated not only on how much it simplifies a system, but on the extent to which that resulting system obeys reasonably reliable (and hopefully simple) laws.

This two-fold criterion---simplification and prediction---extends to the coarse-graining of systems that are non-spatial\index{coarse-graining, non-spatial}, but still topological in nature, meaning that there is some notion of what is ``nearby'', closer or further. A classic example is provided by the science of complex networks. Within this field, an entire industry is devoted towards the problem of community detection and network clustering, which tries to group nodes on the basis of the larger network topology. Nodes that are connected to each other are considered ``nearby'' in some important sense, and the community detection problem amounts to developing innovative ways to summarize these relationships and to group nodes in larger clusters, or communities~\cite{Fortunato201075}. However, this is only part of the problem: when deciding between different community detection algorithms for use on a dynamically evolving system, we should also ask about the extent to which the new coarse-graining obeys reliable dynamical laws (see Ref.~\cite{2014arXiv1409.7403W} and references therein). %Going beyond the network case, it is worth nothing that not all processes of interest take place on a topological space. Many systems lack spatiality altogether---we may, for example, even wish to coarse-grain a computer program.

The laws that obtain for a coarse-grained system are known as \emph{effective theories}~\cite{huggett1995renormalisation,dedeo2011effective}.\index{effective theory} The nature of the lossy compression is dictated by the goal of producing objects that lead to good effective theories that allow for description, prediction, and explanation. When we cluster a high-dimensional dataset, we usually hope to find simplified descriptions of its patterns that provide predictive leverage. If a scientist uses $k$-means, say, her goal is to find a simpler description of the world. Rather than a list of dozens of coordinates, she might find ``three clusters with means $\mu_i$, variance $\sigma_i$''. The process is a success if, for example, a point's membership in a cluster reveals useful or unexpected features of its origin or future development. 

When clustering is ``hard''---\emph{i.e.}, when any particular fine-grained description falls under a single coarse-grained category---it can be represented as a tree, or hierarchy. An evolutionary phylogeny provides a simple example, where distinct species can be grouped on the basis of their common ancestors. Whatever the algorithm---$k$-means, phylogenetic reconstruction, graph clustering, multi-dimensional scaling, latent Dirichlet allocation---the new description is simpler and more compact. It is a form of lossy compression that discards much of the original information and, among other things, makes it impossible to reconstruct the original in all its glory. 

When we go beyond the physical sciences, we should not be surprised to discover that we sometimes wish to coarse-grain by destroying \emph{long}-range order. When we describe texts in a bag-of-words model, for example, we count words but throw away all information about word proximity; the arc of a narrative is lost as the words that appear at the beginning and the end are mixed together in a single probability distribution. Tracking n-grams---pairs, triplets, and $n$-word units---can be considered forms of coarse-graining less destructive than simple bag-of-words; preserving more of the original structure, while dropping longer-range correlations (see Fig.~\ref{text}). Coarse-graining a text through bag-of-words is often, for example, a good first start towards finding out which were most likely written by the same author, or in the same time period, or as the raw material for accounts of cultural dynamics~\cite{rule2015lexical}.
\begin{figure*}
\centering
\begin{mdframed}
\noindent (A) What a piece of work is a man! how noble in reason!
how infinite in faculty! in form and moving how
express and admirable! in action how like an angel!
in apprehension how like a god! the beauty of the
world! the paragon of animals!
\vspace{0.5cm}

(B) how like: 2, how infinite: 1, and admirable: 1, piece of: 1, noble in: 1, the paragon: 1, $\ldots$

\vspace{0.5cm}
(C) how: 5: in: 5: of: 3: a: 3: the: 3: and: 2: like: 2, moving: 1, noble: 1, is: 1, reason: 1, $\ldots$

\end{mdframed}
\caption{Coarse-graining a text. Rather than keep track of full word order (A), we can count occurrences, summarizing the text by its vocabulary (C). Less aggressively, we can summarize the abundance of word pairs (B); 2-grams retain more of the structure of the original text while discarding long-range syntactic order.\label{text}}
\end{figure*}

% Interesting approach, although I am not clear on the relationship between, for example, coarse-graining and generalization, or between coarse-graining and censorship on the basis of retrospective pattern-analysis. Does large-scale abstraction from many instances amount to the same thing as what you define as coarse-graining? It is a big leap from the editorial cleansing of symbolic artifacts to the evolution of various stages of political governance, but I will be interested in following where this leads.  -- Merlin Donald comments

An ideal coarse-graining not only summarizes the full system at any point in time, but provides descriptions with a useful---if probabilistic---logic connecting them together. Much remains to be done in understanding the relationship between how we coarse-grain, and why: the ways in which a particular desire (summary, prediction, explanation, understanding) in a particular field (social, biological, physical) suggests a particular algorithm. 

Rate distortion theory\index{rate-distortion theory} is one of the simplest mathematical accounts, where the loss is quantified in terms of a single utility function that can be understood in terms of an organism's action policy~\cite{marzen2015evolution}. Organisms encode the world in such as way as to minimize dangerous confusions (not mistaking a tiger for a tree) while coarse-graining away irrelevant details (not distinguishing a tiger from a lion). We may well, however, want to go beyond this canonical paradigm to consider coarse-grainings that are predictive, comprehensible, or easy to compute with~\cite{2014arXiv1409.7403W}. This is the domain of machine learning, broadly conceived, and these questions remain at the forefront of the field.

This section has considered the problem of how to coarse-grain, or lossily compress, in an optimal fashion (given constraints such as memory, processing power, risk tolerance, and so forth). But how do individuals---intelligent agents such as humans or the non-human animals---actually coarse-grain their world? How do their brains work when they try, and when they do try, what do they end up doing? Optimal models may provide upper bounds to the correct answers, but this is at heart a problem for cognitive science, neuroscience, and psychology. It is also, as we shall see in the next section, the crucial step needed for us to build our account of major transitions in social order.

\section{Minds in the Loop}

Scientists summarize, but it is not only as dispassionate observers that we attempt to simplify, and thereby predict and understand, our worlds. To navigate the physical world, for example, we (along with other primates) rely on ``folk physics''~\cite{povinelli2000folk}, a reasonably predictive account of the coarse-grained physical world of medium-sized dry goods, where fundamental laws such as the conservation of energy are routinely violated. Similarly for the biological world: when we study informal human reasoning we find a folk biology~\cite{Keil01102013} that includes, among other things, a notion of an \emph{\'{e}lan vital}, or vital force, permeating living things; such a theory can be found in pre-verbal infants~\cite{Setoh01102013}.\index{folk biology}

Physical and biological laws remain constant over the course of an individual's life. Not so for social phenomena, and the (approximate) laws that connect them. In the modern era, new rules of behavior can emerge overnight; in the past, cultural change of this form might have been slower, but still far more rapid than the ten-thousand year timescales of biological evolution.

The fundamental units of social laws are what we might call social facts:\index{social facts} coarse-grained summaries of the beliefs and actions of the vast numbers of people. Without such summarizes in hand, we are lost: we cannot follow norms unless we learn their essence from the behavior of others; we cannot respect authority if we cannot perceive it. We use these coarse-grained summaries to predict and understand the actions and beliefs of others.

Informal examples abound, but one of the clearest quantitative examples can be found in theories of social power. Whether in a modern high school or the banking world of Renaissance Florence, some individuals are perceived to have more power---of the relevant sort---than others. Some bankers are considered more reliable, even if they have little or no capital to back their debts~\cite{padgett2011economic}; some high school students have more power even if their talents and intrinsic charm might argue otherwise~\cite{vaillancourt2003bullying}. 

Power is both created by, and summarizes, the interactions of a society. A vast body of literature in the social sciences has repeatedly returned to this basic phenomenon: how the manifold interactions within a social group lead to hierarchy of status that bears some---but often not very much---relationship to the original intrinsic properties of the individuals themselves~\cite{mann1986sources}. Power thus provides our first explicit example of a socially relevant coarse-graining. To know social power is to know more than just facts about individuals: it is to summarize innumerable facts about the thoughts individuals have about each other, and thoughts about those thoughts, and so forth.

In the modern era, and driven by advances in our studies of non-human behavior, we have come to quantify these hierarchies by \emph{power score}:\index{social power} a single number that summarizes a group consensus on the basis of individual interactions. As they are used in these contemporary studies, power scores compress an $n\times n$ matrix of dyadic interactions to an $n$-element list. There are many individual-level patterns consistent with any particular ranking, but these scores often predict crucial features of an individual's future~\cite{eleanor}, and evolve over time in predictable ways. Extensions of the basic idea---that relative status can be quantified by reference to pairwise interactions---have proven their worth far beyond the academic arena. Among other things, it forms the core of the original algorithms used by Google to summarize collective opinions about the rank-order value of webpages~\cite{pagerank}. These algorithms are fundamentally recursive: to have power is to be seen to have power by those who are themselves powerful.\footnote{Recent work~\cite{eleanor} has distinguished between ``breadth'' and ``depth'' measures of social consensus. Breadth measures measure the power of individual X simply by reference just to the beliefs others have about X. Depth measures, by contrast, also make reference to higher-order facts such as the beliefs others hold about those who hold opinions about X. At least some work has confirmed the greater predictive power of depth measures~\cite{hobson}, providing additional evidence that social facts are not simply compressions of individual-level beliefs, but complex, non-decomposable compressions where every $n(n-1)$ dyadic interaction influences each power score.}

An observer equipped with panoptic and high-resolution data, and an algorithm such as eigenvector centrality, can measure social power. Individuals in the society itself, tasked with the day-to-day problem of decision-making, and operating under biological constraints of both memory and perception, face a much harder task. The models they make of their social worlds must not only strive for accuracy. Models must lead to representations that are intelligible to, and computable by, the agents themselves~\cite{krakauer2010intelligent}. 

In the final analysis, it is the individual who uses these representations to decide what to do. Of course, in doing so, she and her fellows alter the very coarse-grained representations that they rely upon. Understanding the process of belief formation in the presence of an overabundance of information is a key challenge in understanding how the loop between individual behavior and group-level facts is closed.

One of the ways in which individuals collectively understand their social worlds is through the use of novel signaling channels that allow for a collective summarization of a more rapid and complex series of individual-level events. These new signal channels can smooth out irrelevant noise, and make the underlying social patterns visible to the group as a whole. This account, and its supporting empirical evidence, was developed by Refs.~\cite{flack2006encoding,flack2007context,flack2012multiple,flack2013timescales}, with the example of the social construction of power in pigtail macaques.\index{pigtail macaques} Rather than fight, an individual of this species can send a uni-directional subordination signal, ``silent bared teeth'' (SBT), which both inhibits conflict, should it be imminent, and provides information about time averages over past outcomes. The coarse-graining here is over time, summarizing the outcomes of multiple conflicts with a single binary variable. The work of Ref.~\cite{eleanor} ties these same signals to the distributed consensus in the system as a whole, making the coarse-graining over the social network as well.\footnote{The role that SBT plays in primate societies seems to meet the main criteria for what John Searle\index{Searle, John}, in Ref.~\cite{searle2008freedom}, refers to as a status function. SBT is not intrinsically an act of subordination: it does not put the user at an immediate physical disadvantage as, say, similar signals in the canine case. Furthermore, its function is made possible by the collective acceptance of this signal. It allows sender and recipient to avoid conflict in part, presumably, because it is understood as such not only by the pair themselves, but---given the public nature of power and the role of third-party interactions---by the group as a whole. This account of SBT in primate society pushes Searle's (somewhat fanciful) account of the origin of status function a few hundred thousand years further back. The conjectured contextual meaning of the SBT---that it functions, in part, to indicate facts about a pair-wise relationship to third parties---distinguishes it from simpler cases such as that of the alarm call or warning signal.}\index{signal systems}

A study of a different, though still socially complex, species, the monk parakeet~\cite{hobson2014socioecology},\index{monk parakeets} provides another view on how individuals come to know, and act on, coarse-grained facts. Recent collaborative research shows evidence for emergent loop closure in this species as group behavior develops over time. When parakeets first encounter each other during group formation, aggressive behavior appears strategically unstructured. Over time, however, and as individuals become aware of rank order, they appear to direct individual aggressions strategically and based on relative rank. 

High-resolution data on this \emph{knowledge-behavior} loop~\cite{hobson} provides a dynamical picture of how individuals come to know the implicit hierarchies of their world, and alter their behavior in response. In contrast to the pigtail macaques of the example above, monk parakeets appear, so far, to lack a separate signaling system. The density of interactions, however, may allow for participants in this second example to use small, cognitively accessible network motifs to predict the relevant aspects of these coarse-grained power scores. 

Macaques short-circuit violent conflict by signaling social consensus on power; parakeets use the same variables to strategically direct aggression against rank peers. The work of Ref.~\cite{padgett2011economic}, alluded to above, provides an instructive version in the human case drawn from the early years of merchant banking in Renaissance Florence. In the absence of open records, Florentine bankers attempted to reconstruct not only the potential solvency of their colleagues but, crucially, the ideas about that solvency held by others. To know whether someone was a good risk was to know, in part, whether others thought they were. In response to this challenge, bankers, in their letters to each other, summarized facts about their own prestige and solvency, and the prestige and solvency of others, through an elaborate system of rhetoric and telling details that, on the surface, appeared highly tangential to the financial matters at hand~\cite{mclean2007art}. 

When we use machines, in the modern era, to predict features of our society, we often turn, as Google does, to algorithms that rely on successive coarse-grainings of high-resolution data. The recent success of deep learning~\cite{bengio2009learning} is in part due to its ability to adapt, at the same time, its method of coarse-graining and its theory of the logic of those coarse-grained variables. Once we realize that the machine-aided predictors of a system are also participants, it is natural to ask how their use of that knowledge, accurate or not, back-reacts on the society itself. Financial markets provide examples of both positive reinforcement, as in the case of the 2010 Flash Crash~\cite{easley2011microstructure}, and negative reinforcement, as traders destroy the very patterns that provide their source of profit~\cite{Timmermann200415}. We understand very little about how the introduction of these prediction algorithms, on a large scale, will lead to novel feedbacks that affect our political and social worlds; it remains an understudied and entirely open topic.

Whether driven by inference from context or signal, processed by evolved brains or optimized machines, the feedback loop that results from action on the basis of social facts is likely to be a widespread feature of biological complexity. It may extend well beyond the cognitive and even down to molecular scales~\cite{flack2013timescales,walker2013algorithmic}. In the case of interacting individuals, the closure of this loop is a precondition for the causal efficacy of high-level descriptions. It represents our first major transition in political order. Empirical work strongly suggests that this transition happens in the pre-human era. Monkeys, and even parakeets, are quite literally political animals.

\section{Broken Windows and the Normative Pathway}

Defusing a conflict by signal alone, using relative power to adaptively guide aggression, lending to a high-prestige bank: in each of these examples, individuals infer social facts and use them for their own advantage. Some species, however, with humans the most notable example, reason not only from wants and needs, but also according to how they feel things ought to be.

In observing a power structure we may learn new strategies to thrive, but we may also perceive it as just or unjust, legitimate or illegitimate, and these latter perceptions hinge not only on what is and what will be, but on what \emph{should} be. The modal structure of these beliefs is not one of possible worlds, but of deontic logic, how ``things are done'' by ``people like us'' or, in the modern era (as we describe below), how things compare to an ideal standard~\cite{chellas1980modal}. Norms are, in their most developed form, facts about shared ideals, about what the group believes---or, more formally, a coarse-grained representation and lossy compression of the idiosyncratic beliefs and desires held by individuals. We need not all believe exactly the same thing in order to share a norm; norms constitute a new set of group-level facts.\index{social norms}

As with the case of power, facts about norms can not be reduced to the interactions between two individuals. How a norm of politeness works in a particular commercial transaction, for example, depends very little upon what the participants desire. If it is a norm to thank the shopkeeper, a shopkeeper who asks his customer to forego a ``thank you'' may find his request denied or obeyed at best reluctantly; if his counter-normative requests persist, he may find himself shunned by the community as a whole.\footnote{The customer herself may find the request intrinsically unpleasant; norms, once learned, act directly on our emotions. Violations can cause both pain and pleasure, over and above the consequences of the action itself.} To be polite is not to respect someone as they desire to be respected, but to to play out certain patterns of behavior that can reasonably be interpreted as respect in a social context.\footnote{The hypothetical agent that does this interpretation is referred to as the ``big Other'' in some philosophical theories~\cite{sep-lacan,lacan1998seminar}.}

As suggested by the example of just and unjust power, the emergence of a norm can provide a novel pathway for individuals to respond to pre-existing group-level facts. The normative perception of a hierarchy as unjust should be distinguished from the thought that it might be upended for the agent's benefit. We are able to recognize a situation as unjust, and to respond to this injustice emotionally, even when we have no ability to alter it, and even when we might, for other reasons, consider it an injustice necessary on balance.

In humans some normative responses, including the ability to invent a game and play by its rules, seem to be acquired very early in life~\cite{tomasello2009we}. More elaborate norms are learned by observing the community. They are, therefore, predictions: a norm that ceases to have an effect on behavior is unlikely to be so described a few years later. A norm may have an effect without being obeyed---``more honored in the breach than the observance''---but this is exceptional. We can say, with great confidence, that when two men in American society meet to conduct a lengthy business transaction, they will begin by shaking hands.

Yet we use norms for more than prediction. It is unlikely for the weaker player in an unevenly-matched game of tennis to win; it is unlikely for the loser to refuse a handshake at the end. Given knowledge of the strength of the handshaking norm, the responses to these two unlikely events will be distinct. We may re-evaluate our ratings of the two players based on the final score. Yet even if no formal rule requiring a handshake exists, our responses to the norm violation will involve shunning the individual and group-level shaming; examples of how this (rare) violation is discussed in the press confirm the intuition~\cite{cnn2011rahm}.\index{shaking hands}

Norms are critical for the maintenance of social stability, and a long tradition in game theory seeks to describe how altruistic norms may emerge from purely self-interested motives (see, \emph{e.g.}, the critical review of Ref.~\cite{bowles2009microeconomics}), or evolutionary group selection~\cite{akccay2009theory}. In this sense, norms are simply a more elaborate, potentially gene-driven, version of the prudential strategizing described in the previous section. In contrast to individual strategies, however, norms must be shared, and require not just knowledge, but mutual use. Norms play the role of a choreographer that allows multiple individuals to solve joint action problems by coordinating around a specific equilibrium~\cite{gintis2009bounds}; if we do not share the right norm, for example, having access to a punishment mechanism in a public goods game will lead to anti-social, rather than pro-social, results~\cite{herrmann2008antisocial}.

In contrast to lab-based experiments, much of the complexity of ethnographic research comes from the parsing out of the layered and often counter-intuitive roles that norms play in human society. In part due to this complexity, the underlying cognitive mechanisms required for norms to exist and to influence behavior are hotly debated. As reviewed by Ref.~\cite{de2014natural}, reconciliation behaviors (``making up'' after conflict), responses to unequal rewards, and impartial policing may provide examples of non-human normative reasoning. Both reconciliation and responses to inequality are found across multiple taxa. Meanwhile, ``knockout'' studies have verified the causal role of policing~\cite{flack2005robustness,flack2006policing}---if it is understood as a norm, it is a norm that matters. While reconciliation and inequality responses may be understood as negotiated one-on-one norms, policing provides an example of a strictly community-based norm, where individuals attempt to preserve group consensus. 

A separate school of thought, reviewed in Ref.~\cite{tomasello2009cultural,tomasello2014natural}, ties normative behavior to the ability to act on the basis of a belief about what ``we, together, are doing'', the capacity for joint intentionality.\index{joint intentionality} Joint intentionality is often considered a precondition for human society~\cite{searle2008language,searle2010making}; evidence for joint intentionality in non-human animals may come from the example of chimpanzees that engage in group hunting (as opposed to opportunistic, simultaneous chasing)~\cite{call1998distinguishing,bullinger2014chimpanzees}. To require joint intentionality for norm following, however, may set the standard too high, drawing a firm boundary on the basis of cognitive skills where we might expect shades of grey~\cite{andrews2009understanding}.

Rather than drill down to the level of these basic mechanisms, we take a particular example from recent empirical work to look for the distinct traces that normative reasoning leaves on the logic of society as a whole. We do so using a series of investigations into the dynamics of conflict in the editing of Wikipedia. 

Now over fourteen years old, the community surrounding the online encyclop\ae dia has attracted an enormous amount of scholarly attention, both as a laboratory of human interaction and as a phenomenon in its own right~\cite{bar2014twelve}. Ethnographers have studied the culture of Wikipedia editors~\cite{reagle2010good,jemielniak2014common}, finding diverse motivations and self-conceptions among the hundreds of thousands of volunteers who massively outnumber the roughly one-hundred paid employees of the parent foundation.\index{Wikipedia}

Wikipedia is hardly immune to conflict, much of which focuses on article content: what to include in an article and how to represent it. Users who edit pages---particularly, controversial pages associated with political figures such as George W. Bush or Josef Stalin, or conflicts such as Israel--Palestine~\cite{yasseri2014most}---often find they disagree about which facts to include and the prominence those facts should be given. Facts shade naturally into interpretation, and even when all users involved agree on which sources to cite, disagreements do not cease.

Arguments often reduce to competitive editing: one user adds text, a second one modifies it to change the implication, connotation, or weight, the original author, or a new third party, intervenes to shift the tone again. When this process degenerates, and cooperation breaks down completely, editors may resort to what is called a \emph{revert}: completely undoing the work of a previous editor. Reverts are an excellent way to study conflict on large scales because they can be easily identified by machine, rather than by hand-analysis or complex natural-language processing, and we have learned a great deal about collaboration by tracking conflict in this fashion~\cite{yasseri2012dynamics}.

Reverts also have the advantage of being a clear norm violation. Multiple policy pages discuss how one ought not to revert: reverts are described as ``a complete rejection of the work of another editor'' and ``the most common causes of an edit war''; rather than revert each other, editor disagreements ``should be resolved through discussion''; and editors are ``encouraged to work towards establishing consensus, not to have one's own way''. Those whose edits are reverted are urged to turn the other cheek: ``If you make an edit which is good-faith reverted, do not simply reinstate your edit''.\footnote{Drawn from pages current as of 1 April 2015; see {\tt http://en.wikipedia.org/wiki/Wikipedia:Reverting}; {\tt http://en.wikipedia.org/wiki/Wikipedia:Edit\_warring}.} 

Naturally, reality is far more complicated. Reverts are common, in some periods and for some pages rising to nearly half of all edits made. The very fact that the norm is imperfectly obeyed, however, makes it possible to study the dynamics of how users learn and adjust their behavior in response to the actions of others. In empirical study, we find long-range memory intrinsic to periods of inter-revert conflict: the more edits a page has had without a revert, the less likely it is to see a revert on the next edit. Ref.~\cite{dedeoplos} found a two-parameter model for this process, the collective-state model, where the probability of a revert varies as a function of the number of edits, $k$, since the last revert,
\begin{equation}
P_k(R) = \frac{p}{(k+1)^\alpha},
\label{bro}
\end{equation}
where $p$ and $\alpha$ are constants. When $\alpha$ approaches zero, reverts are uncorrelated and conflict arises without regard for context. Over a wide range of pages, however, we find that $\alpha$ clusters around one-half, leading to a simple \emph{square-root law}: the probability of future conflict declines as the square-root of the amount of conflict seen so far. (In this simple model, the clock resets on the appearance of new conflict.) The law appears robust to a wide range of filters, including the inclusion of partial reverts and the restriction to harder conflicts, where we track conflict by a double-revert, \emph{i.e.}, measure the probability of two reverts in a row. The observed timescales of these runs are short, often only hours or even minutes long, requiring us to refer to intrinsic features of the interaction rather than events in the real world, and involve many users, making these interactions intrinsically social, rather than pairwise~\cite{dedeo2014group}.

One can think of Eq.~\ref{bro} as describing a \emph{reverse broken-windows} effect\index{broken windows}. In the original account of broken windows, popularized by Wilson and Kelling~\cite{wilson1982broken}, minor norm violations led to an increasing likelihood of future violations (a single broken window in an abandoned building attracts more). Here, we find the reverse effect: norm-conformant actions lead to an increasing likelihood of future norm conformance.

Based on this result, Ref.~\cite{dedeo2014group} constructed a game-theoretic model to back-infer the underlying beliefs and desires of the users from their behavior alone. In the spirit of earlier work in inductive game theory~\cite{dedeo2010inductive,dedeo2011evidence}, our fundamental goal was to understand the cognitive complexity of the individuals, and how they reacted to the contexts in which they found themselves. We did so using an extensive-form public goods game called the stage game. The stage game models the step-by-step pattern of interaction on a single page, where users interact with those who came before, while setting the stage for the editor who comes next.\index{inductive game theory}

Our analysis of the stage game showed that, under the assumption of a self-reinforcing equilibrium, a very simple model can explain the behavioral data if and when users have context-sensitive utility. Put another way, a parsimonious model is possible when what other people have done in the past not only affects what a user \emph{does}, but what a user \emph{wants}. In order to explain why users edit the way they do, we can not simply describe them as learning how to maximize utility under a fixed tolerance for conflict: we must allow that tolerance to change. Rather than describe a population with a mixture of mutualists and defectors, we have a population whose individuals become mutualists as they see others around them shift towards cooperation themselves.\index{preference change}

This is what we expect if the underlying behavior is truly driven by a normative injunction, where the adherence to the norm by others increases our own desire to conform. On the one hand, we can describe this result in the folk-psychological language of wants and desires. On the other hand, however, we have long known that successful cooperation in public goods games requires mechanisms such as punishment and reputational damage for those who violate norms (see Ref.~\cite{bowles2011cooperative} and references therein). Extensions to those classic results include those of Ref.~\cite{burton2015payoff}, which notes that changing preferences for cooperation can in some cases be explained by individuals learning how they may, or may not, be punished for behaving badly. 

In the language of our model, changing utility functions can represent either shifts in intrinsic desire, or the expectation of future punishment through other pathways. Our inability to split this atom, when the punishment pathway is hidden from view, is a limitation of utility theory itself, which quantifies desires along a single axis. Ref.~\cite{dedeo2014group} found that norm-conformity accumulates faster when individuals interact ($\alpha$ driven towards one), suggesting that reputation drives learning. A ``cheap-talk'' result---norm-conformity is not affected by use of associated discussion pages---further complicates the analysis.

Whether or not this increasing cooperativity is to be referenced to good citizens (changing desires) or good laws (effective incentives)~\cite{bowles2015machiavelli}, we are firmly in the world of norms: patterns of behavior, understood as group-level standards, and enforced by both community action and by individual desire, forced or free. Individuals adjust their behavior in response to what they observe in others; in the example here, simple coarse-grained heuristics on overall levels of cooperation can provide knowledge of the implicit standard. The feedback effects of their responses to this knowledge provide an example of a fundamentally normative form of loop closure, and our second major transition in political order.

\section{Going together to get along: Norm Bundles}

\begin{figure*}
\centering
\includegraphics[width=\textwidth]{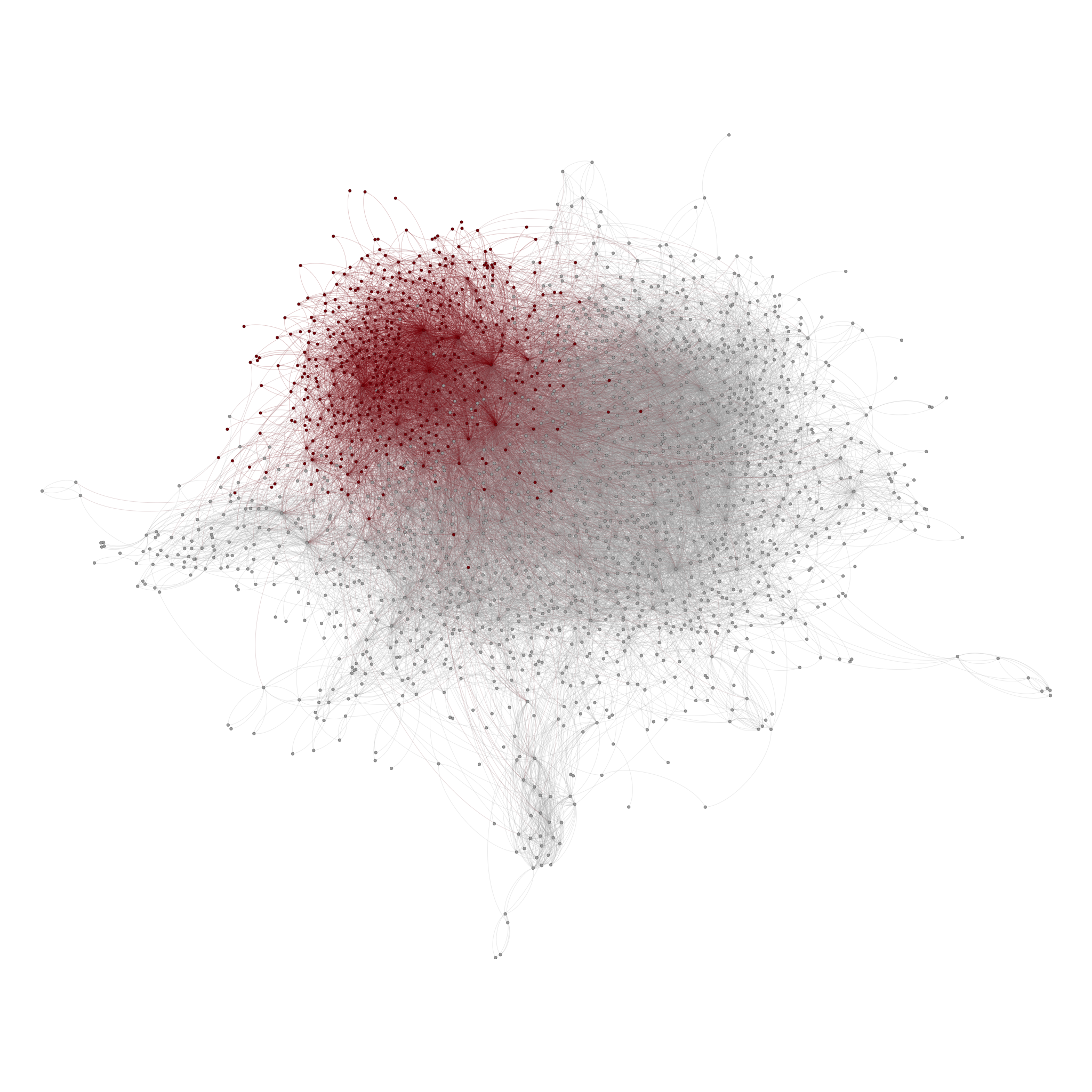}
\caption{Norm bundles on Wikipedia. Nodes refer to policies, guidelines, and essays; links indicate cross-references. Dense clusters of cross-referenced norms range from how to decide whether a person, place, or event is notable enough for inclusion, to how and when to split articles into subtopics, to appropriate and inappropriate ways to handle the stress of online conflict. The largest sub-community is represented as a darker cluster of nodes in the top-left of the network, as found by Louvain clustering~\cite{1742-5468-2008-10-P10008}; this bundle describes norms of article writing, including the need for neutrality (NPOV) and verifiability. Other bundles describe norms of interpersonal interaction such as civility and the assumption of good faith, norms associated with administrative systems, and norms on the use of intellectual property. These top four bundles include just over 75\% of all pages; See Ref.~\cite{bradi}. \label{norm_bundle}}
\end{figure*}

Should Israeli settlements be described as ``key obstacle to a peaceful resolution'', or ``a major issue of contention''? On 2 July 2007, three Wikipedia editors debated these six words on the ``talk'' page of the article on the Israel-Palestinian conflict. Over the next eleven days, the discussion grew to include over twenty editors and ran to over 16,000 words. On July 13th, the last arguments were made, and two of the three original participants had come to agreement on the final wording.

As might be expected, much of the debate centered around the details of the conflict itself, becoming at times only tangentially related to the wording in question. About thirty hours in to the argument, however, the user Jayjg wrote, succinctly, ``WP:NPOV says that opinions cannot be stated as fact, and must be attributed to those who hold them. WP:V says that opinions must be sourced. That should solve the problem; follow policy.'' WP:NPOV is a community abbreviation for a norm that urges editors to adopt a neutral point of view towards article subjects; WP:V, for the norm that all statements in the encyclopedia be verifiable, particularly when challenged by others.

These abbreviations are more than shorthand. In the HTML of Jayjg's comment, they linked to pages in a separate space of the encyclopedia where the two norms are discussed in detail. These two pages are only a small fraction of the nearly 2,000 norm-related pages that users have created over the lifetime of the encyclopedia~\cite{bradi}. Themselves under continual discussion and revision, they have grown to encompass nearly every aspect of the mechanics of article writing, interpersonal interaction, and a small ``administrative'' class given special privileges within the system as a whole.

Those in conflict on Wikipedia may encounter, for example, the norm to assume good faith, often referred by the abbreviation AGF. Users might remind each other of this norm when they believe conflicts are driven by unfair assumptions about the other party. The associated page describing AGF links, among other things, to a (collectively written) essay entitled ``Don't call a spade a spade'' (don't label other users as norm violators; abbreviated NOSPADE); NOSPADE couples, by its own out-link, the AGF norm to both the CIVIL norm (``be respectful and considerate'') and the NPOV norm, urged by Jayjg in his original comment, where NOSPADE violations are likely to occur.\index{civility}

The connections between these norms are not logically necessary: one can imagine a different pattern, where the NPOV norm is supported by a strong (here fictional) PROSPADE norm, with users encouraged to identify and critique each other's underlying motivations. In the Wikipedian bundle, however, AGF, NOSPADE, CIVIL, and NPOV are understood as reinforcing structures that provide coherence to a user's expectations. Given the difficulties of text-based communication, the Wikipedia community choice is likely to be adaptive. 

Not every normative injunction can be uniquely related to core practices; the LONDONDERRY norm, for examples, describes an internal consensus from 2004 on a controversial naming decision. Examples of potentially adaptive clusters abound: Wikipedia's encouragements for users to undertake creative action without interference include networked norms such as OWN (``no-one is the owner of any page''), BUILDER (``don't hope the house will build itself'') and even DHTM (``don't help too much'') and MYOB (``mind your own business''). 

This is an example of a more general principle: once created, norms rarely stay as isolated oughts. We want to make sense of our world and constraints on cognitive load naturally lead to the formation of \emph{norm bundles}..\index{norm bundles} These networks of interacting and self-supporting norms reenforce each other by providing logical or emotive support. One norm is now understood as a natural consequence, or a sub-case, of another. We regularize---\emph{i.e.},  simplify and systematize---in a variety of linguistic and non-linguistic domains~\cite{lieberman2007quantifying,ferdinand2013regularization}. Norm bundling may be driven, as well, by this same instinct to avoid the costs of memory through the systematization of exceptional cases.

In the case of Wikipedia we can build a network from how norms interact, reinforce, and modify each other. We see the emergence of clusters, where basic principles form high-degree cores within distinct communities, and serve as a common point of reference for more peripheral subgraphs; see Fig.~\ref{norm_bundle} for a representation of the full network, as well as the largest sub-bundle, that includes both  NPOV and verifiability.

Wikipedia may be unusual in its ratio of norm to action. It is difficult not to be impressed by the thousands of pages users have created presenting, discussing, and interpreting their community's standards. It may even be worrying: many wiki-like systems appear to fall into a ``policy trap'', where content creation is replaced by policy discussion dominated by a smaller, in-group elite: a modern, electronic version of the Iron Law of Oligarchy~\cite{shaw2014laboratories}. 

Wikipedia is not, however, unusual in the complex ways in which its norms cross-link, how it draws on a set of core principles to carry the periphery along, or how a bundle its may be more than the sum of its parts. An example at the national level is provided by the United States Supreme Court, which in {\it Griswold v.\ Connecticut} (381 U.S.\ 479, 1965) described a right to privacy. This right, nowhere explicitly stated in the Constitution, is the implication of norm bundle, an example of how, in the words of Justice Douglas,  ``specific guarantees in the Bill of Rights have penumbras, formed by emanations from those guarantees that help give them life and substance''. Psychologically, our moral injunctions do not appear to us as statements that we can analyse in isolation, nor even as as directed chains of derivations; they are, instead, dense networks of social practices, mixing rational arguments of greater or lesser plausibility with central emotional, narrative and even mythic appeals~\cite{Cavell1979-CAVTCO,donald1991origins,bellah2011religion}.

\begin{figure*}
\centering
\begin{tabular}{cc}
\includegraphics[width=0.5\textwidth]{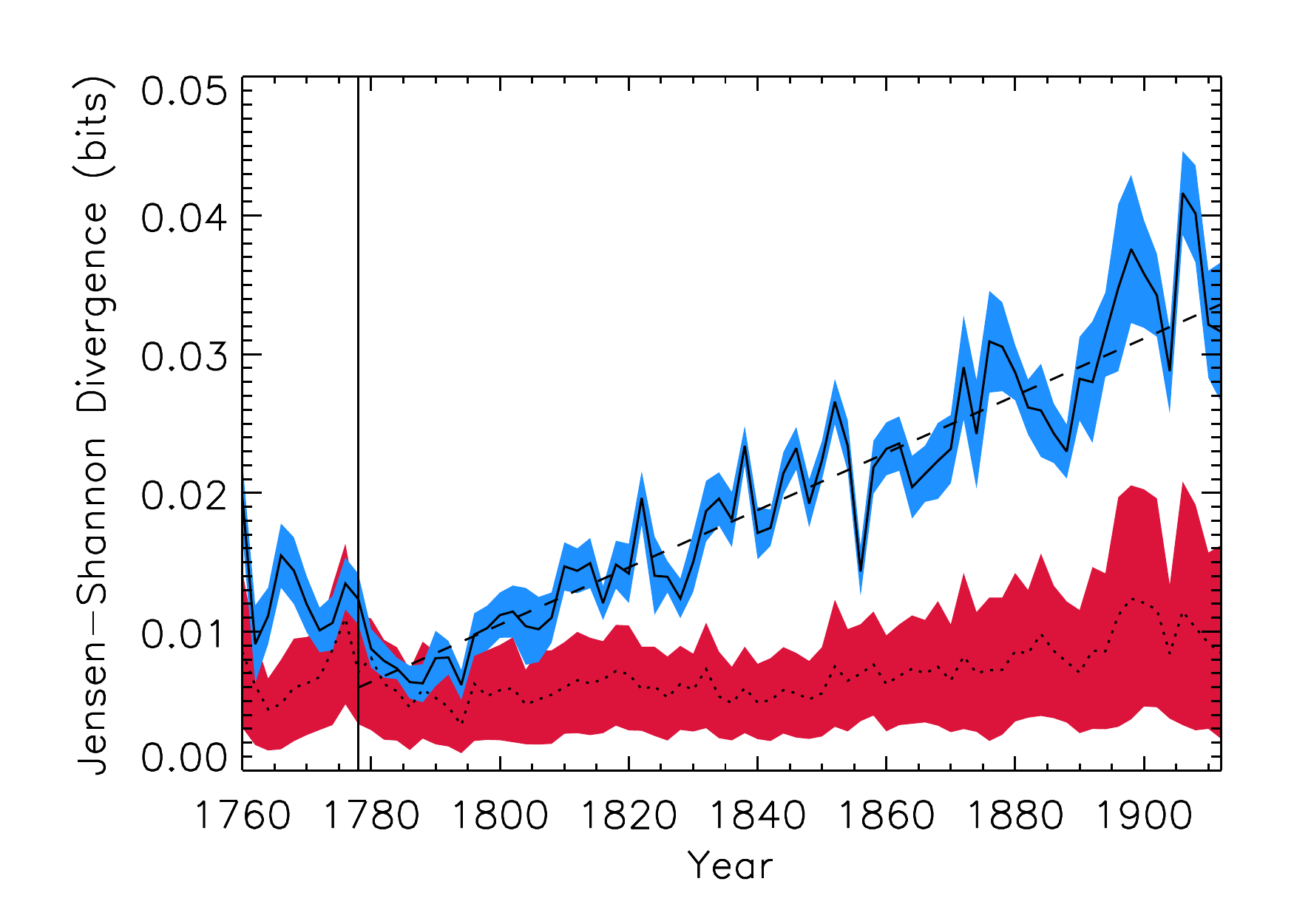} \\ \includegraphics[width=0.5\textwidth]{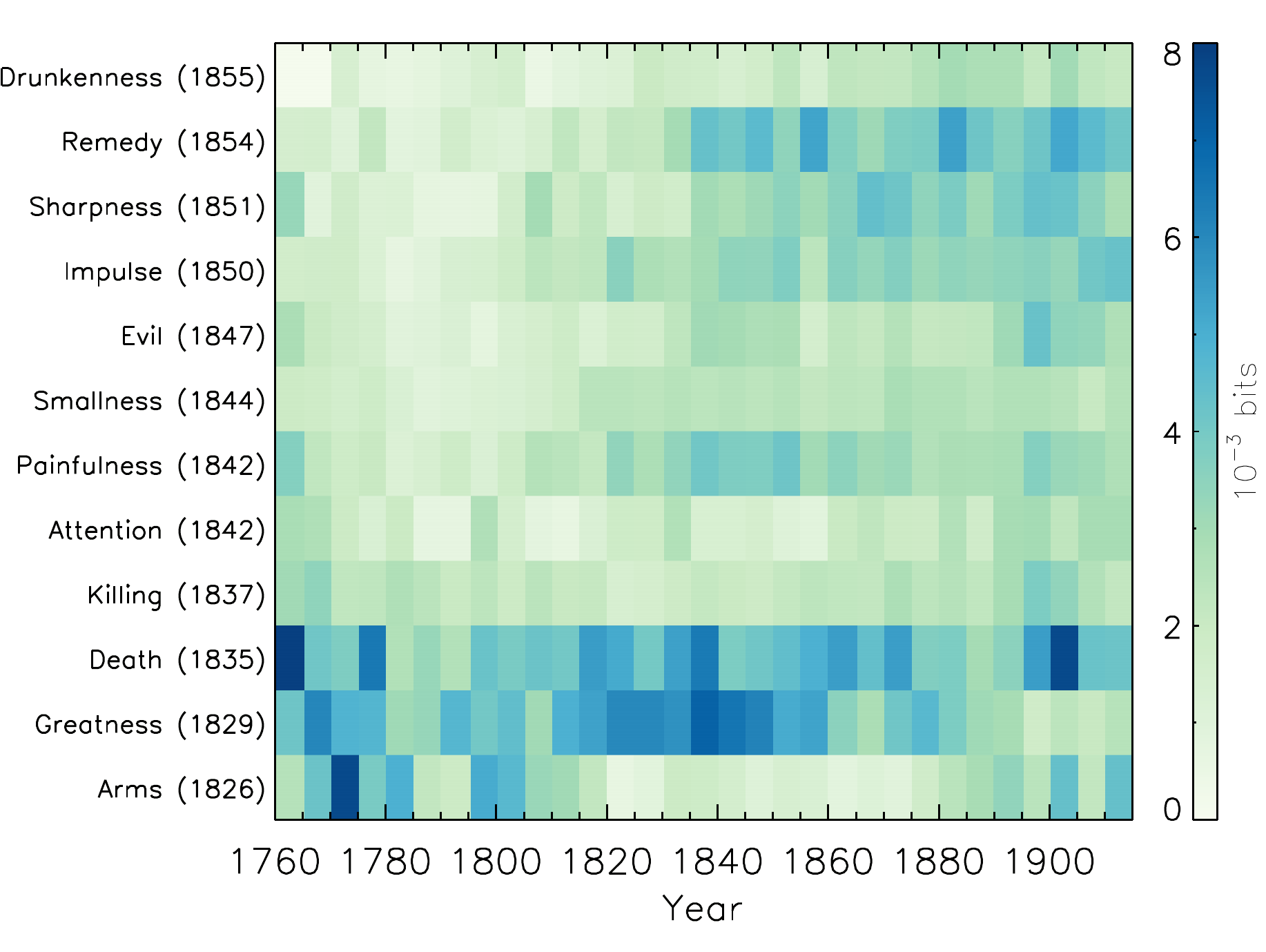}
\end{tabular}
\caption{Correlated norm shifts in discussions of violence, 1760--1913. Top: trials for violent and non-violent offences become increasingly distinct over time, as measured by the Jensen-Shannon distance between spoken text in the two categories (rising blue line, dashed fit). Bottom: the top dozen classes that serve to signal trials for violence. Some, such as references to death, are strong signals of a concern for violence throughout our data. Others, such as those referring to medical evidence (``remedy'') and drunkenness appear much later. Adapted from Ref.~\cite{obo}. \label{norm_shifts}}
\end{figure*}
Norms both interact with, and drive, changing material contexts. Yet because of the reinforcing nature of norm bundles, shifts in behavior are rarely due to the emergence or strengthening of a single norm in isolation. Rather, when studying long term norm-driven change, we expect signals of multiple, conceptually distinct---but bundled---norms working together. We can see this in our analysis of the Old Bailey which began this chapter. In the left-hand panel of Fig.~\ref{norm_shifts}, we show how speech during trials for violent and non-violent crimes became increasingly distinct, tracking the bureaucracy's increasing concern to manage, specifically, the violence of its population~\cite{obo}. This plot tracks the strength of signals, at the one-gram level, that distinguish transcripts describing crimes the court considered violent from those it did not. In the early years of our data, little to no distinction exists; classifications, at least at the one-gram level, appear arbitrary. But from 1780 through to the end of our data in 1913, a long-term secular trend becomes clearly visible, showing how this signal first emerged, and then began to strengthen over time.\index{violence}

These cultural shifts in the attention to violence parallel long-term declines in the homicide rate~\cite{eisner2003long}. The majority of the cases in our data, and the majority of the signal in Fig.~\ref{norm_shifts}, concerns crime less serious than murder. The signal we track is tied to an increasing sensitivity to the ``dark matter'' of violence---the assaults, kidnappings, and violent thefts that do not leave a dead body for demographers to trace.

In the right-hand panel of Fig.~\ref{norm_shifts}, we look closer, into the signal structure itself, to see how the words that signaled these distinctions changed over time. To build signal-to-noise, we group words into synonym sets, so that the set ``impulse'' includes words such as kick, hit, blow and strike; the set ``remedy'' includes words like hospital and doctor; the set ``greatness'', words like very, great, many, much and so; the set ``sharpness'', words like knife, razor and blade.

Studying the changing patterns of these signals gives us clues to the nature of the norm bundles that underlie Britain's transition from the 18th to the 20th Century. Already by 1770, discussions of death were strong signals that the court had indicted the defendants for a violent crime, as were words associated with firearms. However, words such as knife and cut, or hit and strike, took longer to emerge as signals; the case of Junque, Smith, and Leethorp that opens this chapter provides an example of how, early on, assault and the use of a knife, openly discussed before both judge and jury, were able to appear in a trial ostensibly for the non-violent offence of grand larceny.

As the court paid greater attention to more minor forms of violence, parallel shifts occurred in related domains. The sets ``smallness'' and ``attention'', containing words associated with (among other things) observation and measurement also come to prominence: violence must not only be minimized, it must also be measured. Doctors were called upon to provide medical evidence, showing how concerns with lesser forms of aggression led to demands for a scientific account of its effects. In the final decades of our data, words associated with drunkenness emerge, both because the state increasingly attends to the opportunistic violence associated with drinking, and because it is used as an explicit excuse by the defendants themselves: participants attend to violence's external, material causes.

This is what we expect from normative bundling, and a general theory should provide new insight into other phenomena as well. Some norms are extremely adaptive, but many are simply epiphenomenal, like the ritual handshakes of the tennis match. Handshakes can be faked, and are costless forms of cheap talk; norm bundles explain the persistence and pervasiveness of these epiphenomenal norms by reference to the role they play in the larger structure. 

Fig.~\ref{norm_shifts}, by selecting only those topics that contribute to the distinction in question, should not be understood as promoting a Whiggish account~\cite{butterfield1965whig} of norms in concert combining to produce the modern world. Norms within a bundle do not always work in the same direction, and we expect frustration and disagreement. Incipient conflict can be been seen in a graph theoretic analysis of the Wikipedian bundles shown in Fig.~\ref{norm_bundle}, where norms encouraging users to ``ignore all rules'' (IAR) in seeking creative ways to improve the encyclopedia maintain a large topological distance from norms specifying, in microscopic detail, conventions for transliterating Belarusian (BELARUSIANNAMES). The NPOV norm links, among other things, to pages describing how to resolve naming conflicts, but also to a user essay entitled ``civil POV pushing'', describing concerns about users who, through persistence and careful adherence to interpersonal norms such as AGF and CIVIL, tilt pages in ways that violate NPOV.

A more serious example of intra-bundle conflict, in the case of the common law, can be found in the doctrine of felony murder, where courts punish people for the unintended consequences of a crime. A death caused by the negligent, but accidental, destruction of a traffic signal may be treated as a civil matter. Conversely, a teenager who steals a stop sign and thereby causes a fatal accident may be tried for manslaughter.\footnote{An example of the former is Dixie Drive it Yourself System v.~American Beverage Co. (Louisiana Supreme Court; 1962), where a negligent driver knocked over signal flags leading to a fatal accident. An example of the latter is the State of Florida v.~Christopher Cole, Nissa Baillie and Thomas Miller (1997) where the three defendants received 15 year sentences for a (confessed) stop-sign theft that, two or three weeks later, led to a fatal accident. Review of the Florida case focused on whether it was that stop sign in particular that had been stolen, if too much time had elapsed, and on inappropriate behavior by the prosecutor---not on the fundamental linking of the theft and unintended death.} The general principle, that one can be punished for an unintended consequence of a conceptually distinct crime, is a sufficiently ancient part of common law that as early as 1716 it was treated as a self-evident fact~\cite{hawkins1824treatise}. It persists in the United States today, but is widely seen, elsewhere, to be in fundamental conflict with co-bundled injunctions against strict liability and in favor of the need for \emph{mens rea}~\cite{binder2007culpability}.

Over a decade ago, Ehrlich and Levin~\cite{levin} posed a series of questions for scientists interested in the emergence of norms. Referring to the ``regrettably infertile'' notion of the meme, they urged renewed attention to the development of theories to both quantify and explain the process of cultural evolution. A decade later, the 23 questions they posed remain unresolved---despite massive progress in the development of meme-contagion models and game theoretic accounts of multi-agent interaction. Many of their questions focused on individual-level cognition, including the origin of novel ideas in a mind, the decision to adopt ideas from others, and the covariance of these cognitive processes with other facts about an individual.\index{meme}\index{contagion model}

If our account is correct, we can make new progress by combining the multi-agent approach common to both contagion and game-theoretic models with the cognitive questions of Ehrlich and Levin. Studies of individual level cognition, however, must be used for more than simply fixing the parameters of an agent-based model. Groups, not individuals, construct norm bundles; and individuals that must then learn them both from direct inspection, but also from watching how others behave and extrapolating a mental representation that may differ a great deal from the massively complex structure of Fig.~\ref{norm_bundle}.

We understand little of what is required for norm bundling to begin. If non-human animals have norms, do they have bundles? Is norm bundling a gradual transition, as groups begin by pairing norms, or do large bundles emerge suddenly, at a critical point? Gradual or sudden, the emergence of norm bundling represents a new level of complexity in how individuals perceive, and respond to, their social worlds. It provides our third example of a major transition in political order.

\section{Conclusions}

The most ambitious theories of cultural evolution extend into biological time. When they do so, they often divide history into epochs marked by dramatic shifts in cognitive complexity~\cite{donald1991origins,bellah2012axial_merlin}. Drawing on this tradition, we have focused on transitions in the causal pathways between group-level facts and the individual. When minds are in the loop, coarse-graining is no longer just a method for understanding the material world. It becomes a constitutive part of what it means for a collection of individuals to become a society. If this is correct, the origins of society may share a great deal of their causal structure with the origins of life itself~\cite{walker2013algorithmic}.

Reference to the capacities of the individual mind is a common theme in political theory, and the {\it Leviathan} of Thomas Hobbes\index{Hobbes, Thomas} opened, in 1651, with a theory of cognitive science. Once we recognize the importance of the feedback loop between the individual and the group, however, it becomes harder to distinguish between changes in an individual's ability, and the social scaffolding necessary to support it~\cite{zawidzki2013mindshaping}. Are Wikipedian norms supported by coordinated punishments and rewards that manipulate simple, self-interested utility maximizers? Or do they involve a desire to conform, pride in one's reasonableness, or the notion of an ideal standard for an electronic public sphere? It is hard to imagine an ideal that everyone holds but no-one rewards or enforces. Yet it is hard, also, to imagine people shunning and shaming, praising and rewarding, without adopting the norms themselves and with an eye solely on the causal outcomes of each individual act; the cognitive burden is too high.%\footnote{It may also be insufficient. Punishments must be supported by second-order punishments for those who fail to punish or do so inappropriately. Second-order punishments require, of course, a third system of sanctions, and the tower extends. \emph{Quis custodes iposos custodes}.} The minds of individuals may be so intertwined, and their capacities so dependent on context and communication, that the original distinction has no scientific ground. 

In the past, mathematical theories of social behavior have oversimplified both the human mind and the societies it creates. To counter that tendency we have, in this chapter, attempted to provide vivid portraits of some of the systems under study. Social worlds, like biological ones, are intrinsically messy. They build themselves through bricolage, constantly repurposing small details for new ends~\cite{levi1966savage,balkin2002cultural}. Details abound, may later come to matter, and should be respected: at the very least, we expect their statistical properties will play a role in future mathematical accounts.

Conversely, to mathematize a problem is to allow its examples to be compared across context and scale. If we understand the ecological rationality~\cite{gigerenzer1999fast} of signaling systems that naturally coarse-grain noisy mechanisms, we may find common explanations for how signals work across culture and species. If we study the fine-grained dynamics of cooperation in a contemporary system, we may be able to reverse-engineer how cultures in the past bundled norms to govern the commons. If we can build a network theory of the emergence of norm bundles, we may be able to compare vastly different societies to find common patterns in cultural evolution.

Deep histories of social complexity~\cite{smail2007deep,fukuyama2011origins} are often narratives. This does not mean, however, that quantitative accounts must be restricted to system-specific studies. Laws of social dynamics are expected to be probabilistic, but they may be laws nonetheless, able to accurately describe, explain, and even predict the world at a particular resolution. Our work here suggests that these dynamics are driven in part by top-down causation and complex feedbacks between the individual and the group. It suggests a new role for interdisciplinary collaboration between the cognitive and social sciences. And the complexity of these systems suggests a critical role for the interpretive scholarship of political theorists, ethnographers, and historians.

\section*{Acknowledgements}

I am grateful to audiences at the Interacting Minds Center of Aarhus University, Denmark, the Ostrom Workshop in Political Theory and Policy Analysis of Indiana University, the Global Brain Institute of Vrije Universiteit Brussel, Belgium, and the Santa Fe Institute, where early versions of this work were presented. I thank Merlin Donald, Tim Hitchcock, Dan Smail, Colin Allen, Jerry Sabloff, John Miller, Alexander Barron, and Natalie Elliot for readings of this work in draft form. This work was supported in part by National Science Foundation Grant \#EF-1137929, by a Santa Fe Institute Omidyar Fellowship, and by the Emergent Institutions project.

\newpage
\printindex
\newpage

\bibliographystyle{unsrt}
\bibliography{new_draft}

\end{document}